\documentclass[prb,twocolumn,superscriptaddress,nofootinbib]{revtex4-2}

\usepackage{multirow}%
\usepackage[T1]{fontenc}
\usepackage{xcolor}
\usepackage{pdfcolmk}
\usepackage{mathrsfs}
\usepackage{amsmath}
\usepackage{amssymb}
\usepackage{graphicx}
\usepackage{wasysym}
\PassOptionsToPackage{normalem}{ulem}
\usepackage{ulem}
\usepackage[
unicode=true,pdfusetitle,
bookmarks=true,bookmarksnumbered=false,bookmarksopen=false,
breaklinks=false,pdfborder={0 0 0},pdfborderstyle={},
backref=false,colorlinks=true,
hypertexnames=false
]{hyperref}
\hypersetup{citecolor=blue,urlcolor=blue}

\makeatletter
\providecolor{lyxadded}{rgb}{0,0,1}
\providecolor{lyxdeleted}{rgb}{1,0,0}

\DeclareRobustCommand{\lyxsout}[1]{\ifx\\#1\else\sout{#1}\fi}

\usepackage{bbm}
\usepackage[final]{changes}
\usepackage{ulem}

\begin{document}

\title{Landscape index and its boundary correspondence in inhomogeneous chiral systems}

\author{Guo-Jian Qiao}
\thanks{These authors contributed equally to this work.}
\affiliation{Graduate School of China Academy of Engineering Physics, Beijing 100193, China}

\author{Zhi-Lei Zhang}
\thanks{These authors contributed equally to this work.}
\affiliation{Graduate School of China Academy of Engineering Physics, Beijing 100193, China}

\author{C. P. Sun}
\email{suncp@gscaep.ac.cn}
\affiliation{Graduate School of China Academy of Engineering Physics, Beijing 100193, China}

\begin{abstract}
We define a real-space topological index from the geometry
of a quadratic landscape associated with the spatial dynamics,
which is governed by the zero-energy eigenvalue equation of
chiral systems. The landscape index remains invariant under
continuous deformations preserving the uniform exponential
separation between decaying and growing spatial solutions
and equals the continuum winding number in the homogeneous
limit. Across an inhomogeneous interface, the difference
between the indices on the two sides equals the protected
net chiral zero-mode count, without requiring asymptotically
homogeneous bulk regions. Explicit landscape constructions
yield analytical bounds on Hamiltonian parameters and their
spatial gradients that determine the indices for all profiles
satisfying these bounds, without specifying their detailed
spatial form. Dimensional reduction extends the landscape
index and its boundary correspondence to higher-dimensional
systems. The resulting bounds guarantee protected radial
zero modes in effective Kitaev-type systems of arbitrary
transverse codimension and identify parameter and momentum
regions supporting chiral edge states in periodically
modulated quantum anomalous Hall systems. For the quasiperiodic interface considered, the index
difference guarantees a single protected chiral branch despite persistent modulation on both sides, as confirmed by numerical calculations. Landscape geometry thus connects real-space topology to the protected net chirality of boundary states.
\end{abstract}

\maketitle

\section{Introduction}
In one-dimensional chiral systems with translational symmetry, topological phases are characterized by a
winding number defined in momentum space~\cite{Ryu2010}. This invariant remains unchanged under continuous
deformations that preserve the bulk gap and, through the bulk--boundary correspondence, determines
topologically protected boundary states~\cite{Chiu2016}. Spatial inhomogeneity generally removes translational symmetry, so the conventional Bloch winding number is no longer directly available.

Real-space topological invariants extend this topological characterization to disordered
chiral systems~\cite{Prodan2016,Mondragon2014,Loring2017,Lin2021}. Scattering-matrix formulations determine topological invariants from reflection matrices at zero energy~\cite{Beenakker2011,Fulga2011}, while transfer matrices and Lyapunov exponents relate the edge index to the spatial decay of zero-energy states~\cite{Graf2018}. These formulations generally extract topological information from a specified real-space Hamiltonian. When only bounds on spatially varying parameters are known, a complementary goal is to determine the protected net chiral count for every profile compatible with those
bounds.

The landscape approach introduced in our previous
work~\cite{qiao2026landscape} addresses the existence
and counting of boundary zero modes directly through
bounds on Hamiltonian parameters and their spatial gradients. In this approach, the zero-energy equation is
recast as a first-order spatial dynamical system, with the spatial
coordinate serving as the evolution parameter. A quadratic landscape is constructed as a Lyapunov
function for this spatial evolution~\cite{Lyapunov1992,pnas_Ao,Wang2008,xu2026wave}. The geometry of this landscape, characterized by
the numbers of positive and negative eigenvalues
of its Hermitian matrix, determines the dimensions of the
decaying and growing solution spaces, respectively,
while the intersection of the decaying subspace
with the physical boundary subspace determines
the boundary zero-mode count. This geometric counting relation raises a further
question: can landscape geometry also define an
intrinsic real-space topological index whose
difference across an inhomogeneous interface
determines the protected net chiral zero-mode count?

In this paper, we define such an index from the landscape geometry.
The landscape index is independent of the choice of admissible
landscape and remains invariant under continuous deformations
preserving a uniform exponential dichotomy, which separates
exponentially decaying and growing spatial solutions.
The required dichotomy can itself be established through an
auxiliary landscape satisfying a strict Lyapunov condition,
allowing the dynamical assumption underlying the index to be
verified self-consistently.

In the homogeneous limit, the index equals the continuum winding
number, establishing the landscape--bulk correspondence.
At an inhomogeneous interface, the difference between the two
landscape indices equals the protected net chiral zero-mode
count, establishing the landscape--boundary correspondence
without requiring asymptotic homogeneity on either side.
For an explicit landscape, the inertia and monotonicity conditions
yield analytical bounds on Hamiltonian parameters and their
spatial gradients. These bounds fix the indices for all compatible profiles without
specifying their detailed spatial form or solving the eigenvalue
equation.

We further extend the landscape index and its boundary correspondence to higher-dimensional systems through
dimensional reduction. For effective radial Kitaev-type systems with a radially varying
chemical potential, a norm-preserving transformation yields the
same one-dimensional chiral equation for arbitrary transverse
codimension. A common landscape construction therefore yields analytical
bounds on the chemical potential and its radial gradient
that guarantee a protected zero mode at the hard-wall
inner boundary.

In quantum anomalous Hall systems, the bounds identify parameter
and momentum regions supporting protected chiral edge states
under periodic mass modulation. The channel indices obtained from spatial Floquet analysis agree with these analytical predictions. For the quasiperiodic interface, tail indices $0$ and $1$
imply a single protected negative-chirality interface branch
with linear dispersion within an analytically guaranteed
momentum interval. 
Numerical spectra and probability densities show the predicted
dispersion and interface localization despite persistent
modulation on both sides.

\section{Landscape index and its invariance}
We consider a one-dimensional Hamiltonian with chiral symmetry $\{\Gamma,\mathcal H\}=0$ and
$\Gamma^2=1$. In the eigenbasis of $\Gamma$,
the Hamiltonian takes the off-diagonal
form~\cite{Ryu2010}
\begin{equation}
    \mathcal H=
    \begin{bmatrix}
        0&q\\
        q^\dagger&0
    \end{bmatrix}.
    \label{eq:offdiag}
\end{equation}
Writing $\Psi=(\Psi_+,\Psi_-)^T$, where
$\Psi_\pm$ are the components in the
$\Gamma=\pm1$ subspaces, the zero-energy
equation $\mathcal H\Psi=0$ separates into
$q\Psi_-=0$ and $q^\dagger\Psi_+=0$.

In the coordinate representation, let $q$ be a
$p$th-order differential operator acting on
$N$-component wave functions, with a nonsingular
highest-derivative coefficient matrix.
The equation $q\Psi_-=0$ can then be recast as a
first-order spatial dynamical system
$\partial_x\mathbf\Phi=\mathbf M(x)\mathbf\Phi$
of dimension $D=pN$. Here, $x$ serves as the spatial evolution parameter,
and $\mathbf\Phi=[\Psi_-,\partial_x\Psi_-,\ldots,\partial_x^{p-1}\Psi_-]^T$ collects the wave-function components and their
spatial derivatives up to order $p-1$. 
The matrix $\mathbf M(x)$ is determined by the
zero-energy equation and is assumed to be continuous
and uniformly bounded on $[0,\infty)$.

We assume that the spatial dynamics admits
a uniform exponential dichotomy on $[0,\infty)$,
which gives a splitting of the initial-value space
at $x=0$,
$\mathbb C^D=\mathcal E^s\oplus\mathcal E^u$
~\cite{Coppel1978,Sacker1978}.
Initial conditions in $\mathcal E^s$ generate
exponentially decaying solutions, whereas those
in $\mathcal E^u$ generate exponentially growing
solutions under forward spatial evolution. To characterize this decomposition
geometrically, we introduce the Hermitian
quadratic landscape
$\mathcal L=\frac12\mathbf{\Phi}^\dagger
\mathbf G\mathbf{\Phi}$. A landscape is admissible if $\mathbf G(x)$ is
continuously differentiable and Hermitian, both
$\mathbf G$ and $\mathbf G^{-1}$ are uniformly
bounded, and $\partial_x\mathcal L\leq0$ along
every spatial trajectory. The monotonicity condition is equivalent to
$\partial_x\mathbf G+\mathbf G\mathbf M
+\mathbf M^\dagger\mathbf G\leq0$. Since $\mathbf G(x)$ is continuous and nonsingular, its inertia $\operatorname{In}\mathbf G(x)=(n_+,n_-,0)$ is independent of $x$,
where $n_+$ and $n_-$ count its positive and
negative eigenvalues, respectively.
The landscape theorem~\cite{qiao2026landscape}
then gives $\dim\mathcal E^s=n_+$ and
$\dim\mathcal E^u=n_-$.
We therefore define the landscape index as
\begin{equation}
    \nu_{\mathcal L}
    \equiv
    \frac{n_+-n_-}{2}
    =
    \frac{\dim\mathcal E^s-\dim\mathcal E^u}{2}.
    \label{eq:nul-def}
\end{equation}

The first equality in Eq.~\eqref{eq:nul-def}
defines the landscape index through the geometry
of the landscape, characterized by the numbers
of positive and negative eigenvalues of $\mathbf G$. At a fixed $x$, the nonempty regions where
$\mathcal L$ is positive or negative are
homotopy equivalent to spheres whose dimensions
determine the corresponding eigenvalue counts.
Thus, this geometry encodes topological information. 
The second equality in Eq.~\eqref{eq:nul-def}
relates the landscape index to the dimensions
of the stable and unstable subspaces of the
zero-energy spatial dynamics.
Since these dimensions are intrinsic to the
spatial dynamics, the landscape index is independent
of both the reference position and the choice
of admissible landscape. A detailed derivation of the correspondence between the landscape inertia and the dimensions of the
stable and unstable subspaces, together with proofs
of these homotopy equivalences, is given in
Appendix~\ref{sec:landscape-inertia}.

We next establish the invariance of the landscape
index under continuous deformations of the Hamiltonian.
Consider a family of chiral-symmetric Hamiltonians
$\mathcal H_\lambda$, with $\lambda\in[0,1]$.
At zero energy, the corresponding spatial dynamics
is described by
$\partial_x\mathbf\Phi_\lambda
=\mathbf M_\lambda(x)\mathbf\Phi_\lambda$.
The differential order $p$ and spatial-system
dimension $D$ are fixed throughout the deformation.
We assume that the uniform exponential dichotomy
is preserved for every $\lambda$ and that
$\mathbf M_\lambda$ depends continuously on
$\lambda$ in the uniform norm:
\begin{equation}
    \lim_{\lambda\to\lambda_0}
    \|\mathbf M_\lambda
    -\mathbf M_{\lambda_0}\|_\infty=0,
\end{equation}
where
$\|\mathbf M\|_\infty
=\sup_{x\geq0}\|\mathbf M(x)\|$.

The roughness theorem for exponential dichotomies
guarantees that sufficiently small perturbations
of $\mathbf M_\lambda$ in the uniform norm preserve
the exponential dichotomy and the dimensions of
the stable and unstable
subspaces~\cite{Coppel1978,Ju2001}.
Consequently,
$\dim\mathcal E_\lambda^{s,u}
=\dim\mathcal E_{\lambda_0}^{s,u}$
for $\lambda$ sufficiently close to $\lambda_0$.
Although the subspaces themselves may vary,
their dimensions are locally constant.
Since $[0,1]$ is connected, these dimensions
remain unchanged throughout the deformation.
It follows from Eq.~\eqref{eq:nul-def} that
\begin{equation}
    \nu_{\mathcal L}(\lambda)
    =
    \nu_{\mathcal L}(0),
    \qquad \lambda\in[0,1].
    \label{eq:homotopy}
\end{equation} 
Thus, the landscape index is protected
against sufficiently small perturbations
and remains invariant under continuous
deformations preserving the uniform
exponential dichotomy. This is the
real-space analogue of momentum-space topological invariants
under gap-preserving deformations. An index change along a continuous deformation at fixed $D$ therefore requires a breakdown
of the uniform exponential dichotomy. The proof of this
homotopy invariance is given in detail in Appendix~\ref{sec:landscape-inertia}.

The exponential dichotomy assumed above can itself be established through the landscape construction. Under the stated boundedness assumptions, the strict Lyapunov condition
\begin{equation}
    \partial_x\mathbf G+\mathbf G\mathbf M
    +\mathbf M^\dagger\mathbf G
    \leq-c\mathbb I_D,
    \qquad c>0,
\end{equation}
uniformly for $x\geq0$ ensures a uniform exponential dichotomy~\cite{Mitropolsky2002}. If the original landscape satisfies only the nonstrict monotonicity condition, it suffices to construct an auxiliary landscape with the same inertia, whose matrix and inverse are uniformly bounded and which satisfies this strict condition. This auxiliary landscape construction thus verifies self-consistently the dynamical condition required for the landscape index to be well defined~\cite{qiao2026landscape}.

\subsection{Landscape--bulk correspondence}

We now relate the landscape index to the
bulk topological invariant in the homogeneous
limit. For $\mathbf M(x)=\mathbf M$, we seek
exponential solutions
$\mathbf\Phi(x)=e^{izx}\mathbf\Phi_0$,
where $z=k+i\kappa$ with $k,\kappa\in\mathbb R$.
For the zero-energy equation $q\Psi_-=0$,
substituting $-i\partial_x\to z$ gives
$q(z)\Psi_{-,0}=0$. A nontrivial solution
therefore requires
\begin{equation}
    Q(z)\equiv\det q(z)=0.
    \label{eq:Qzero}
\end{equation}
Each root $z$ corresponds to a spatial
eigenvalue $iz$ of $\mathbf M$.
Since $e^{izx}=e^{ikx}e^{-\kappa x}$,
roots in the upper and lower half-planes
correspond to exponentially decaying and
growing solutions, respectively.
This correspondence also holds for repeated
roots when their algebraic multiplicities
are counted. Consequently, the numbers
$N_+$ and $N_-$ of roots in the upper
and lower half-planes satisfy
\begin{equation}
    N_+=\dim\mathcal E^s,
    \qquad
    N_-=\dim\mathcal E^u.
    \label{eq:roots-subspaces}
\end{equation}

For real $z=k$, $q(z)$ reduces to the
momentum-space block $q(k)$ of the homogeneous
Hamiltonian. A gapped bulk requires
$Q(k)=\det q(k)\neq0$ for all real $k$. The continuum winding number~\cite{Ryu2010,Lee2024} is given by
\begin{equation}
    \nu_c
    =
    \frac{1}{2\pi i}
    \int_{-\infty}^{\infty}dk\,
    \partial_k\ln Q(k).
    \label{eq:winding}
\end{equation}
The $p$th-order operator $q$ has a nonsingular highest-derivative coefficient matrix, so $Q(z)$ is a polynomial of degree $D=pN$. Closing the integration contour in the upper half-plane, the argument principle relates the contour integral to the number $N_+$ of enclosed zeros. Since the leading term of $Q(z)$ is proportional to $z^D$, the semicircle at infinity contributes $D/2$ to the contour integral normalized by $1/(2\pi i)$. The integral along the real axis therefore gives $\nu_c=N_+-D/2$.
Using $N_+ + N_-=D$ and
Eq.~\eqref{eq:roots-subspaces}, we obtain
\begin{equation}
    \nu_c
    =
    \frac{N_+-N_-}{2}
    =
    \frac{\dim\mathcal E^s-\dim\mathcal E^u}{2}
    =
    \nu_{\mathcal L}.
    \label{eq:bulk-correspondence}
\end{equation}

Equation~\eqref{eq:bulk-correspondence}
establishes the landscape--bulk correspondence:
the landscape index equals the continuum
winding number of the homogeneous Hamiltonian.
Thus, bulk topology is encoded in the
geometry of the landscape.
Details of the root counting and contour
derivation are given in Appendix~\ref{app:roots_subspaces}.

The root correspondence also relates the breakdown
of the exponential dichotomy to bulk gap closing.
For constant $\mathbf M$, the exponential
dichotomy breaks down precisely when a
spatial eigenvalue $iz$ reaches the imaginary
axis. Since $\operatorname{Re}(iz)=-\kappa$,
this occurs when $\kappa=0$, giving
\begin{equation}
    Q(k)=\det q(k)=0,
    \label{eq:gapclosing}
\end{equation}
which is the condition for bulk gap closing.
The breakdown of a uniform exponential dichotomy
can therefore be regarded as a real-space
generalization of the bulk-gap-closing condition.
Under a uniform exponential dichotomy, the landscape index is well defined and remains invariant under continuous deformations that preserve this dichotomy.

To compare relative bulk indices, consider two
homogeneous bulk Hamiltonians, denoted by $L$
and $R$, with the same spatial-system dimension
$D$.
Let $\nu_{\mathcal L}^{L,R}$ and $\nu_c^{L,R}$
denote their landscape indices and continuum
winding numbers, respectively, and let
$N_+^{L,R}$ count the zeros of $Q^{L,R}(z)$
in the upper half-plane, including algebraic
multiplicities.
Applying the landscape--bulk correspondence
to each bulk gives
$\nu_{\mathcal L}^{L,R}
=\nu_c^{L,R}=N_+^{L,R}-D/2$.
The common term $-D/2$ therefore cancels
in their difference, yielding
\begin{equation}
    \nu_{\mathcal L}^{R}-\nu_{\mathcal L}^{L}
    =
    \nu_c^{R}-\nu_c^{L}
    =
    N_+^{R}-N_+^{L}
    \in\mathbb Z.
    \label{eq:bulk-difference}
\end{equation}
Thus, the difference between the landscape indices equals the difference between the bulk winding numbers and is integer-valued even when the individual indices are half-integers. The next section establishes that this difference determines the net chiral zero-mode count at an interface, without requiring asymptotic homogeneity on either side.

\subsection{Landscape--boundary correspondence}

We assume that the zero-energy spatial
dynamics admits uniform exponential
dichotomies on both half-lines, with
admissible landscapes on each side.
An interface zero mode must decay toward
both spatial infinities.
For $q\Psi_-=0$, let
$\mathcal E_L^u(0)$ and $\mathcal E_R^s(0)$
denote the subspaces at $x=0$ corresponding
to solutions decaying as $x\to-\infty$
and $x\to+\infty$, respectively.
A zero mode therefore requires
$\mathbf\Phi(0)\in
\mathcal E_L^u(0)\cap\mathcal E_R^s(0)$,
which ensures both spatial decay and
matching at the interface.
Since $\mathbf\Phi(0)$ uniquely determines
the solution of the spatial-evolution equation,
the number of linearly independent
negative-chirality zero modes is
$\mathcal N_-=\dim(\mathcal E_L^u\cap\mathcal E_R^s)$.

For the adjoint equation
$q^\dagger\Psi_+=0$, positive-chirality
zero modes must likewise decay toward
both spatial infinities.
The decaying subspaces of the adjoint
equation are dual to those of $q$.
Consequently, the number of positive-chirality
zero modes is
$\mathcal N_+
=D-\dim(\mathcal E_L^u+\mathcal E_R^s)$,
which counts the independent directions
not spanned by the two subspaces.
The duality relation and its conditions are established in Appendix~\ref{SM:Fredholm}.

\begin{table}[t]
\centering
\caption{Comparison of momentum-space topology and real-space landscape topology. The landscape formulation applies to spatially inhomogeneous systems without requiring asymptotic homogeneity. In the homogeneous limit, the landscape index equals the continuum winding number.}
\label{tab:topological-correspondence}
\footnotesize
\setlength{\tabcolsep}{4pt}
\renewcommand{\arraystretch}{1.25}
\resizebox{\columnwidth}{!}{%
\begin{tabular}{lll}
\hline\hline
 & Momentum space & Real space \\
\hline
Formulation
& Bloch Hamiltonian $\mathcal H(k)$
& Spatial dynamics $\mathbf M(x)$ \\
Topological index
& Winding number $\nu_c$
& Landscape index $\nu_{\mathcal L}$ \\
Transition
& Bulk gap closing
& Dichotomy breakdown \\
Net chiral count
& $\Delta\mathcal N=\nu_c^R-\nu_c^L$
& $\Delta\mathcal N=\nu_{\mathcal L}^R-\nu_{\mathcal L}^L$ \\
\hline\hline
\end{tabular}
}
\end{table}

The net chiral zero-mode count is
$\Delta\mathcal N
\equiv\mathcal N_- -\mathcal N_+$.
Using the identity $\dim(U\cap V)+\dim(U+V)
=\dim U+\dim V$ for two subspaces
$U$ and $V$, we obtain 
\begin{equation}
    \Delta\mathcal N
    =
    \dim\mathcal E_L^u
    +
    \dim\mathcal E_R^s-D.
    \label{eq:interface-count}
\end{equation}
Thus, although the number of zero modes
in each chiral sector depends on how the
decaying subspaces intersect, their net
count depends only on the dimensions of
these subspaces.
From the landscape inertia theorem,
$\dim\mathcal E_L^u
=D/2-\nu_{\mathcal L}^L$ and
$\dim\mathcal E_R^s
=D/2+\nu_{\mathcal L}^R$.
Substituting these relations into
Eq.~\eqref{eq:interface-count} yields
\begin{equation}
    \Delta\mathcal N
    =
    \nu_{\mathcal L}^R
    -
    \nu_{\mathcal L}^L.
    \label{eq:landscape-interface}
\end{equation}

Equation~\eqref{eq:landscape-interface}
establishes the landscape--boundary
correspondence across an interface:
the difference between the landscape
indices on the two sides equals the
net chiral zero-mode count.
Consequently, the total number of
interface zero modes satisfies
$N_{\rm zero}\geq
|\nu_{\mathcal L}^R-\nu_{\mathcal L}^L|$.
Additional zero modes of opposite
chiralities may couple and shift away
from zero energy under perturbations
preserving chiral symmetry, whereas
the net chiral count remains invariant
as long as the required exponential
dichotomies are preserved.

The net chiral zero-mode count also equals the conventional Fredholm index $\operatorname{ind}q=\dim\ker q-\dim\ker q^\dagger$~\cite{Graf2018}. The landscape index difference therefore coincides with this index:
\begin{equation}
    \operatorname{ind}q
    =
    \nu_{\mathcal L}^R
    -
    \nu_{\mathcal L}^L.
    \label{eqindq}
\end{equation}
When landscapes on both sides can be constructed explicitly from the Hamiltonian parameters, their inertia and monotonicity conditions yield analytical bounds on these parameters and their spatial gradients. Within the resulting parameter regions, the landscape index difference determines the protected net chiral zero-mode count without explicitly solving the eigenvalue problem.

Importantly, the landscape--boundary correspondence remains valid in spatially inhomogeneous systems without requiring
asymptotically homogeneous bulk regions.
When the bulk regions on both sides
become homogeneous and remain gapped
far from the interface, the
landscape--bulk correspondence gives
\begin{equation}
    \Delta\mathcal N
    =
    \nu_{\mathcal L}^R-\nu_{\mathcal L}^L
    =\nu_c^R-\nu_c^L,
    \label{eq:full-bulk-boundary}
\end{equation}
recovering the conventional bulk--boundary correspondence.

Thus, the landscape index reproduces
the bulk winding number in the
homogeneous limit. As a real-space
topological invariant, it not only determines the net chiral zero-mode count but also remains
invariant under continuous deformations
that preserve the exponential dichotomy.
These properties allow the landscape
index to define topologically nontrivial
and trivial phases directly in real
space, even in spatially inhomogeneous
systems. The corresponding momentum-space
and real-space relations are summarized
in Table~\ref{tab:topological-correspondence}.

\section{Dimensional reduction}

The landscape index and landscape--boundary correspondence extend to
higher-dimensional systems that decompose into effective one-dimensional
channels. For a channel labeled by $\eta$, suppose its eigenvalue problem
can be mapped onto the zero-energy equation of an effective one-dimensional
chiral Hamiltonian. The chiral block equation $q_\eta\Psi_-=0$ then defines
the spatial dynamics $\partial_u\mathbf\Phi=\mathbf M(u;\eta)\mathbf\Phi$,
where $u$ is the spatial-evolution coordinate. When the required exponential
dichotomy and landscape conditions hold, the channel index is
\begin{equation}
    \nu_{\mathcal L}(\eta)
    =
    \frac{1}{2}
    \left[
        \dim\mathcal E^s(\eta)
        -\dim\mathcal E^u(\eta)
    \right].
    \label{eq:channel-resolved-index}
\end{equation}
The landscape--boundary correspondence relates this index to boundary
zero modes subject to the physical boundary conditions. At an interface,
$\nu_{\mathcal L}^{R}(\eta)-\nu_{\mathcal L}^{L}(\eta)$ equals the net
chiral zero-mode count. These zero modes belong to the effective chiral
Hamiltonian; their energies in the original system follow from the
channel mapping.

We illustrate this reduction first for radial zero modes in Kitaev-type
systems and then for edge states in quantum anomalous Hall systems with
a spatially modulated Dirac mass. An angular-momentum decomposition,
when available, yields channels along the radial coordinate $r$.
For a planar boundary, conserved parallel momentum
$\mathbf k_{\parallel}$ instead labels channels along the normal
coordinate $x$.

\subsection{Radial landscape in inhomogeneous Kitaev-type systems}
\label{sec:radial}

We first consider an effective radial chiral Hamiltonian with a spatially
varying chemical potential \cite{},
\begin{equation}
    H_d
    =
    -\left[\frac{\hbar^2}{2m}\mathcal D_r^2+\mu(r)\right]\tau_z
    -i\Delta\mathcal D_r\tau_x,
    \label{eq:radial-Hd}
\end{equation}
where $\mathcal D_r=\partial_r+(d-1)/(2r)$. Here, $\mu(r)$ is a smooth
chemical-potential profile confined to a finite window away from zero,
$m$ the effective mass, $\Delta>0$ the pairing strength, and $d$ the
transverse codimension. The Pauli matrices $\tau_i$ act in particle--hole
space. For $d=1$, the model reduces to the continuum Kitaev
chain~\cite{Kitaev2001}, while $d=2$ can describe a selected
angular-momentum channel of a half-flux vortex in a spinless $p$-wave
superconductor~\cite{Read2000,Shen2012}.

On $r\geq r_c>0$, we impose the hard-wall condition $\Psi(r_c)=0$ and
require normalizability with the radial measure $r^{d-1}dr$.
The inner boundary represents the wire end for $d=1$ or an idealized
boundary around the vortex core for $d=2$. The transformation
$\Psi(r)=r^{-(d-1)/2}\widetilde\Psi(r)$ preserves the norm, absorbs the
radial measure, and reduces $\mathcal D_r$ to $\partial_r$.
The transformed wave function is square-integrable with respect to $dr$
and satisfies $\widetilde\Psi(r_c)=0$. The resulting Hamiltonian is
\begin{equation}
    H_{\rm rad}
    =
    -\left[\frac{\hbar^2}{2m}\partial_r^2+\mu(r)\right]\tau_z
    -i\Delta\partial_r\tau_x,
    \label{SM:radial-reduced}
\end{equation}
with chiral symmetry $\{H_{\rm rad},\tau_y\}=0$.

At zero energy, the two chirality sectors decouple. In the
negative-chirality sector, write $\widetilde\Psi(r)=\phi(r)\chi_-$,
where $\chi_-$ is a constant normalized spinor satisfying
$\tau_y\chi_-=-\chi_-$. The radial amplitude obeys
\begin{equation}
    \frac{\hbar^2}{2m}\phi''(r)
    +\Delta\phi'(r)
    +\mu(r)\phi(r)=0,
    \label{SM:radial-scalar}
\end{equation}
where primes denote radial derivatives. Introducing
$\mathbf\Phi=(\phi,\phi')^T$, this equation becomes
\begin{equation}
    \partial_r\mathbf\Phi
    =
    \begin{bmatrix}
        0&1\\
        -2m\mu(r)/\hbar^2&-2m\Delta/\hbar^2
    \end{bmatrix}
    \mathbf\Phi.
    \label{SM:radial-spatial}
\end{equation}
The radial boundary problem therefore takes the spatial dynamical form
introduced above.

For this system, we construct the quadratic landscape
\begin{equation}
    \mathcal L_{\rm rad}
    =\frac12\mathbf\Phi^\dagger\mathbf G(r)\mathbf\Phi,
    \label{SM:radial-inhom-landscape}
\end{equation}
with $\mathbf G(r)=\operatorname{diag}
[\operatorname{sgn}\mu,\hbar^2/(2m|\mu|)]$.
Here, $\operatorname{sgn}\mu=\pm1$ denotes the fixed sign of the chemical
potential. This choice makes the coefficient of $|\phi|^2$ constant
and cancels the mixed terms proportional to $\phi^*\phi'$ in the
spatial derivative. Using Eq.~\eqref{SM:radial-scalar}, we obtain
\begin{equation}
    \partial_r\mathcal L_{\rm rad}
    =
    -\frac{1}{|\mu(r)|}
    \left[
        \Delta+\frac{\hbar^2}{4m}\partial_r\ln|\mu(r)|
    \right]|\phi'|^2.
    \label{SM:radial-inhom-derivative}
\end{equation}
The landscape is nonincreasing with $r$ when
$\partial_r\ln|\mu(r)|>-4m\Delta/\hbar^2$.
For the opposite boundary orientation, choosing the opposite chirality
and reversing the spatial coordinate changes the sign of the logarithmic
chemical-potential gradient. The corresponding condition is
$\partial_r\ln|\mu(r)|<4m\Delta/\hbar^2$.
A stronger uniform sufficient condition valid for either boundary
orientation is
\begin{equation}
    \sup_{r\geq r_c}
    \left|\partial_r\ln|\mu(r)|\right|
    <\frac{4m\Delta}{\hbar^2}.
    \label{eq:radial-gradient}
\end{equation}

With monotonicity ensured by this bound, the radial index follows from
the landscape inertia. The matrix $\mathbf G$ has inertia $(2,0,0)$
for $\mu>0$ and $(1,1,0)$ for $\mu<0$, yielding
\begin{equation}
    \nu_{\mathcal L}^{\rm rad}
    =
    \begin{cases}
        1,&\mu(r)>0,\\
        0,&\mu(r)<0.
    \end{cases}
    \label{eq:radial-landscape-index}
\end{equation}
These values identify nontrivial and trivial phases, respectively.
Under the hard-wall condition at $r=r_c$, the landscape--boundary
correspondence gives one protected radial zero mode in the nontrivial
phase: the negative-chirality sector has two independent decaying
solutions, and $\phi(r_c)=0$ leaves one independent linear combination.
In the trivial phase, the radial index vanishes, and no net chiral zero
mode is protected.

The strict bound in Eq.~\eqref{eq:radial-gradient} leaves a uniform
positive margin in the monotonicity condition. Together with the stated
bounds on $\mu(r)$, it permits an auxiliary landscape matrix with the
same inertia satisfying the strict Lyapunov
condition~\cite{qiao2026landscape}. This establishes the required uniform
exponential dichotomy and completes the self-consistent justification
of the index and boundary-mode count.

These results require no asymptotic homogeneity of $\mu(r)$.
Since the transformed spatial dynamics is independent of $d$, the same
construction applies to these effective radial channels with the
specified inner boundary. In the homogeneous limit, the index recovers
the Kitaev-type classification; at $\mu=0$, a spatial exponent vanishes
and the exponential dichotomy breaks down.

\subsection{Landscape index in modulated QAH systems}

We next consider a two-dimensional quantum anomalous Hall (QAH) system
with a spatially varying Dirac mass $m=m(x)$
~\cite{Liu2008,Rui2010,Klinovaja2015,Martino2023}.
For a boundary normal to $x$, translational symmetry along $y$ conserves
$k_y$. The Hamiltonian reads
\begin{equation}
    \mathcal H(k_y)
    =
    -iA\sigma_x\partial_x
    +Ak_y\sigma_y
    +[m(x)+B(k_y^2-\partial_x^2)]\sigma_z,
    \label{eq:qah-model}
\end{equation}
where $A,B>0$ set the linear and quadratic dispersion scales,
respectively, and $\sigma_i$ are Pauli matrices acting in the two-band
space. For each fixed $k_y$, separating the term $Ak_y\sigma_y$ defines
the transverse Hamiltonian
\begin{equation}
    \mathcal H_x(k_y)=\mathcal H(k_y)-Ak_y\sigma_y,
\end{equation}
which has chiral symmetry, $\{\mathcal H_x(k_y),\sigma_y\}=0$.
A zero mode of $\mathcal H_x(k_y)$ with negative chirality therefore
corresponds to a state of the full Hamiltonian with dispersion
$E(k_y)=-Ak_y$. For a half-plane $x>0$, imposing $\Psi(0)=0$ selects
the physical edge states; at an interface, the state must instead
decay toward both spatial infinities.

Writing $\Psi(x)=\phi(x)\chi_-$ with $\sigma_y\chi_-=-\chi_-$,
the zero-energy equation of $\mathcal H_x(k_y)$ reduces to
\begin{equation}
    \phi''(x)+\frac{A}{B}\phi'(x)
    +\mu(x;k_y)\phi(x)=0,
\end{equation}
where primes denote derivatives with respect to $x$ and
$\mu(x;k_y)=-[m(x)/B+k_y^2]$. This has the same form as the radial
equation. On a region where the effective parameter remains within a
finite window away from zero, choose
$\mathbf G=\operatorname{diag}[\operatorname{sgn}\mu,1/|\mu|]$
for $\mathbf\Phi=(\phi,\phi')^T$. The landscape is nonincreasing under
the uniform sufficient condition
\begin{equation}
    \sup_x
    \left|\partial_x\ln|\mu(x;k_y)|\right|
    <\frac{2A}{B}.
    \label{SM:QAH-gradient-condition}
\end{equation}
As in the radial problem, this strict bound also permits an auxiliary
landscape that establishes the required exponential dichotomy.
The landscape inertia then gives the channel index
\begin{equation}
    \nu_{\mathcal L}(k_y)
    =
    \begin{cases}
        1,&\mu(x;k_y)>0,\\
        0,&\mu(x;k_y)<0.
    \end{cases}
\end{equation}
For the stated hard-wall boundary, a channel with
$\nu_{\mathcal L}=1$ supports one protected chiral edge state.
In the homogeneous limit $m(x)=m$, a negative mass supports this state
within $|k_y|<\sqrt{-m/B}$, whereas a positive mass has
$\nu_{\mathcal L}=0$ for every momentum channel.

To examine spatial modulation, consider
$m(x)=m_0+\delta m\sin(x/\ell)$, where $\delta m>0$ is the amplitude
and $2\pi\ell$ is the period. The landscape conditions yield
\begin{equation}
    \nu_{\mathcal L}(k_y)
    =
    \begin{cases}
        1,
        &k_y^2<-\dfrac{m_0+\delta m C}{B},\\[2mm]
        0,
        &k_y^2>-\dfrac{m_0-\delta m C}{B},
    \end{cases}
    \label{eq:qah-index}
\end{equation}
where $C=\sqrt{1+B^2/(4A^2\ell^2)}$.
For $m_0<-\delta m C$, each momentum channel in the first region
supports one protected chiral edge state. For $m_0>\delta m C$,
all momentum channels have zero index. Equation~\eqref{eq:qah-index} gives sufficient
analytical bounds on the modulation parameters
and $k_y$ for an entire family of sinusoidal
mass profiles. 

\begin{figure}[t]
\centering
\includegraphics[width=8.6cm]{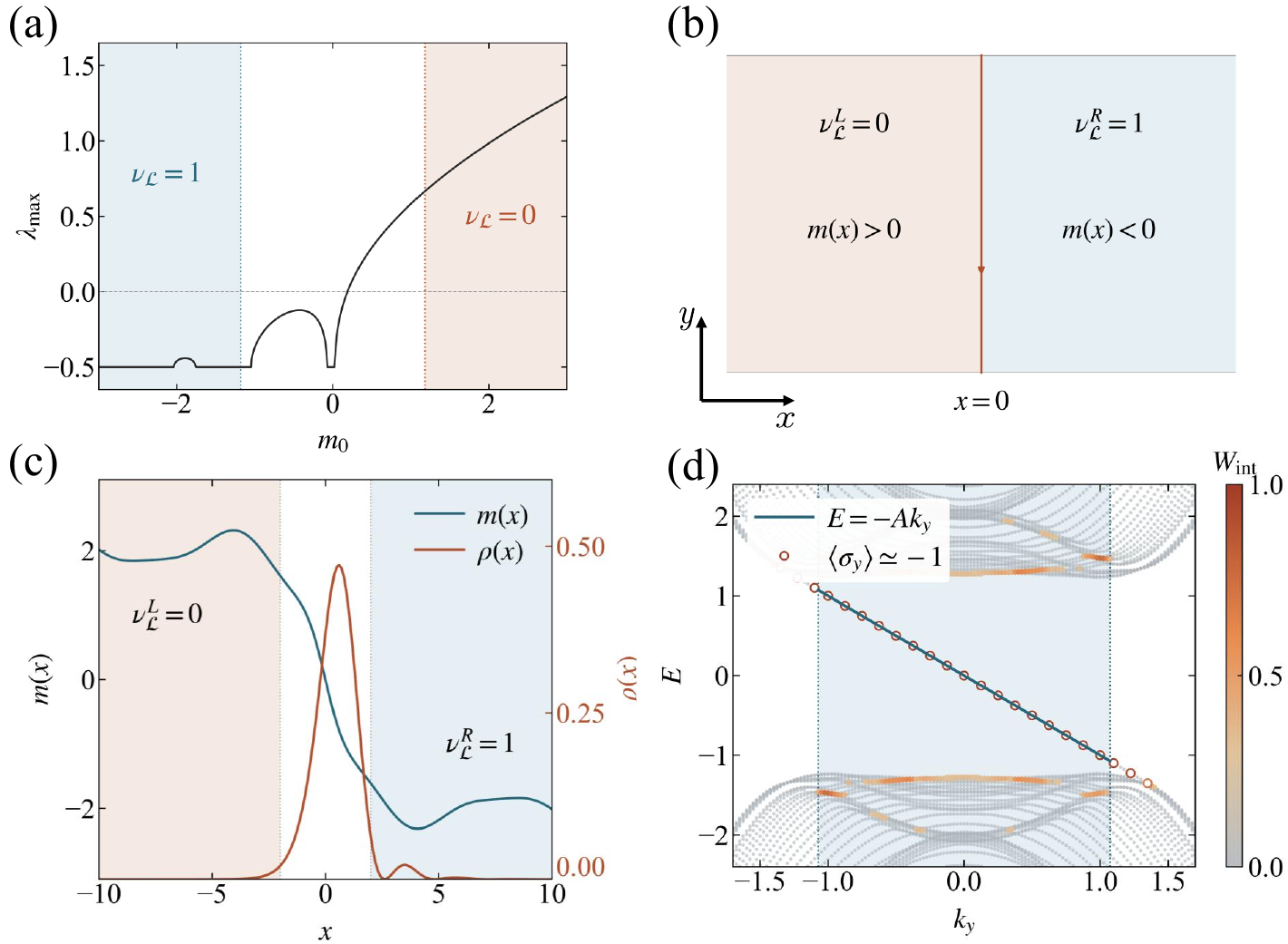}
\caption{Landscape indices and boundary states in a spatially modulated QAH system. (a) Largest spatial Floquet exponent $\lambda_{\max}=T^{-1}\max_j\ln|\rho_j|$ for $m(x)=m_0+\sin(x/0.8)$ at $k_y=0$, where $T=1.6\pi$ and $\rho_j$ are the eigenvalues of the one-period evolution matrix. The dotted lines mark $m_0=\pm C$; the shaded regions have $\nu_{\mathcal L}=1$ (left) and $0$ (right). (b) Interface schematic; the two tails have channel indices $\nu_{\mathcal L}^{L}=0$ and $\nu_{\mathcal L}^{R}=1$. The arrow shows the direction in which the chiral state propagates along the interface.  (c) Mass profile of Eq.~\eqref{eq:qp} (blue) and normalized interface-state density $\rho(x)$ at $k_y=0$ (orange). The shaded tails satisfy the analytical bounds for the indicated indices. (d) Low-energy spectrum obtained by diagonalizing the full Hamiltonian for the profile in (c). Color indicates the interface weight $W_{\rm int}=\int_{-4}^{4}\|\Psi(x)\|^2\,dx$. The interface branch has $\langle\sigma_y\rangle\simeq-1$ and follows $E=-Ak_y$ (solid line). The blue band marks the sufficient interval $|k_y|<1.073$. Lengths and energies are measured in units of $B/A$ and $A^2/B$, respectively, so that $A=B=1$. The numerical spectrum is calculated for a strip with hard-wall boundaries at $x=\pm40$.}
\label{fig:results}
\end{figure}

For the numerical comparisons, lengths and energies are measured in
units of $B/A$ and $A^2/B$, respectively, so that the dimensionless
coefficients are $A=B=1$.
Figure~\ref{fig:results}(a) compares the analytical bounds with the
largest spatial Floquet exponent $\lambda_{\max}$ at $k_y=0$,
using $\delta m=1$ and $\ell=0.8$.
The Floquet exponents are obtained from spatial evolution over one
modulation period and measure exponential growth or decay over
successive periods. For $m_0<-\delta m C$, both exponents are
negative, so both independent solutions decay and
$\nu_{\mathcal L}=1$.
For $m_0>\delta m C$, the largest exponent is positive; since the two
exponents sum to $-A/B<0$, one solution grows and the other decays,
giving $\nu_{\mathcal L}=0$.
The numerical indices agree with the analytical predictions in both
shaded regions. The analytical bounds and
Floquet calculation are detailed in
Appendix~\ref{SM:dimensional-reduction}.

We next consider an interface between two spatially
inhomogeneous regions.
For general $A,B>0$, any smooth bounded mass profile
satisfying $m(x)\geq m_*$ on the left tail,
$m(x)\leq-m_*$ on the right tail, and
$|m'(x)|\leq b_*$ on both tails has landscape indices
$\nu_{\mathcal L}^{L}=0$ and $\nu_{\mathcal L}^{R}=1$ whenever $k_y^2<m_*/B-b_*/(2A)$, where
$m_*>0$ and $b_*\geq0$.
The landscape--boundary correspondence therefore
guarantees one protected chiral interface branch
for every profile satisfying these bounds,
without constraining the detailed mass profile
in the finite interface region.

As an explicit example, consider an interface whose two
sides remain quasiperiodically modulated and do not
approach homogeneous bulk regions
[Fig.~\ref{fig:results}(b)].
In the dimensionless units $A=B=1$, we choose
\begin{equation}
    m(x)=-2\tanh x+0.3\sin x+0.15\sin(\sqrt2\,x).
    \label{eq:qp}
\end{equation}
The first term connects a positive-mass background on the left to a
negative-mass background on the right, while the two incommensurate
sinusoidal terms maintain quasiperiodic modulation arbitrarily far
from the interface [blue curve in Fig.~\ref{fig:results}(c)].

For $x\leq-2$, the mass satisfies $m(x)\geq m_*$, whereas for $x\geq2$
it satisfies $m(x)\leq-m_*$. The gradient obeys
$|m'(x)|\leq b_*$ on both tails. Here, $m_*$ and $b_*$ are obtained
by bounding the three terms in $m(x)$ and their derivatives separately:
\begin{align}
    m_*&=2\tanh2-0.45\simeq1.4781,\notag\\
    b_*&=2\operatorname{sech}^2(2)+0.3+0.15\sqrt2
    \simeq0.6534.
    \label{eq:tails}
\end{align}
For $k_y^2<m_*$, the effective parameter
$\mu(x;k_y)=-[m(x)+k_y^2]$ is negative on the left tail and positive
on the right. On both tails, these estimates imply
$|\mu|\geq m_*-k_y^2$ and $|\partial_x\mu|\leq b_*$.
Consequently, $|\partial_x\ln|\mu||\leq b_*/(m_*-k_y^2)$,
so the landscape condition holds with a uniform positive margin
whenever $k_y^2<m_*-b_*/2\simeq1.1513$.

Within this interval, the tail indices are
$\nu_{\mathcal L}^{L}=0$ and $\nu_{\mathcal L}^{R}=1$.
The decaying subspaces can be propagated through the finite interface
region without changing their dimensions. The landscape index difference
fixes the net chiral zero-mode count to one.
Moreover, $\dim\mathcal E_R^s=2$ and $\dim\mathcal E_L^u=1$,
so the intersection and adjoint counting relations give the exact counts
$\mathcal N_-=1$ and $\mathcal N_+=0$.
The landscape--boundary correspondence therefore guarantees one
protected negative-chirality interface branch with dispersion
$E(k_y)=-Ak_y$ throughout this momentum interval.

Figure~\ref{fig:results}(d) shows the low-energy spectrum obtained by
numerical diagonalization of the full Hamiltonian on a finite interval.
The colors indicate the probability weight within $|x|\leq4$,
identifying states localized near the interface.
Throughout the analytically guaranteed interval
$|k_y|<\sqrt{m_*-b_*/2}\simeq1.073$, marked by the blue band,
the interface branch follows $E(k_y)=-Ak_y$ and has
$\langle\sigma_y\rangle\simeq-1$, in agreement with the predicted
dispersion and chirality. Figure~\ref{fig:results}(c) shows the normalized
probability density at $k_y=0$, confirming
localization near the interface. The numerical branch extends beyond the guaranteed interval, consistent with the sufficient nature of the analytical bounds.

Together, these examples show how dimensional reduction extends the
landscape index and landscape--boundary correspondence to
higher-dimensional systems through effective one-dimensional channels.
The landscape construction yields analytical bounds on Hamiltonian
parameters and their spatial gradients under which the channel indices
fix the protected net chiral counts for all spatial profiles satisfying
these bounds, without specifying their detailed form or solving the
eigenvalue equations.

\section{Conclusion}

We have defined a real-space topological index from the
geometry of a quadratic landscape associated with the
zero-energy spatial dynamics of chiral systems.
The index is independent of the choice of admissible
landscape and remains invariant under continuous
deformations preserving the uniform exponential separation
between decaying and growing spatial solutions.
In the homogeneous limit, it equals the continuum winding
number, establishing the landscape--bulk correspondence.
At an inhomogeneous interface, the difference between the
landscape indices on the two sides equals the protected
net chiral zero-mode count, establishing the
landscape--boundary correspondence without requiring
asymptotic homogeneity on either side.

Explicit landscape constructions yield analytical bounds
on Hamiltonian parameters and their spatial gradients.
These bounds determine the indices and protected net
chiral counts for all spatial profiles satisfying them,
without specifying their detailed spatial form or solving
the eigenvalue equation.
The required exponential dichotomy can itself be established
through an auxiliary landscape satisfying a strict Lyapunov
condition, allowing the dynamical assumption underlying
the index and counting relations to be verified
self-consistently.

Dimensional reduction extends these results to effective
one-dimensional channels of higher-dimensional systems.
For effective radial Kitaev-type systems, bounds on the
chemical potential and its radial gradient guarantee
protected zero modes at a hard-wall inner boundary for
arbitrary transverse codimension.
For quantum anomalous Hall systems, the analytical bounds
identify parameter and momentum regions supporting
protected chiral edge states under periodic mass modulation,
with the predicted channel indices agreeing with spatial
Floquet analysis.
For the quasiperiodic interface considered, the tail index
difference guarantees a single protected chiral branch
within an analytically determined momentum interval,
despite persistent modulation on both sides.
Numerical spectra and probability densities agree with
its predicted dispersion and localization. 

The radial formulation further motivates applications
to topological insulator--superconductor
heterostructures~\cite{Fu_2008_SC_Insulator,Qi2010,Yue_2023,yue2026chiralitybreak}
and semiconductor nanowire--superconductor heterostructures
with radial inhomogeneity~\cite{Volovik1999,Sau_2010_Generic_New_Platform,Reeg_2018_Metallization_nw}.
Landscape geometry thus connects real-space topology to the protected net chirality of boundary states,
while explicit landscape constructions yield analytical
bounds on Hamiltonian parameters and their spatial
gradients that fix the landscape indices and protected
net chiral zero-mode counts.

\begin{acknowledgments}The authors thank Tan-Ji Zhou and Kang Xu at GSCAEP, and Xin Yue at CSRC for helpful discussions. This study is supported by the National Natural Science Foundation of China (Grant Nos. 12688201 and 12547124), the China Postdoctoral Science Foundation (Grant Nos. 2025M784438 and GZC20261845), and the Science Challenge Project (Grant No. TZ2025017).

\end{acknowledgments}
\section*{Data availability}
The data and calculation scripts used for Fig.~\ref{fig:results} are available from the authors upon request.

\appendix
\section{Landscape geometry and topological invariance}
\label{sec:landscape-inertia}

We first establish the correspondence between
the landscape inertia and the dimensions of
the stable and unstable subspaces of the
zero-energy spatial dynamics.
We then demonstrate that the landscape
geometry itself encodes topological information
through the homotopy properties of its
positive and negative regions.
Finally, we prove that the landscape index
remains invariant under continuous deformations
preserving the uniform exponential dichotomy.

\subsection{Stable-unstable subspaces and landscape inertia}

The zero-energy eigenvalue equation $q\Psi_-=0$
can be recast as a first-order spatial dynamical
system
\begin{equation}
    \partial_x\mathbf{\Phi}
    =
    \mathbf M(x)\mathbf{\Phi},
    \qquad
    \mathbf{\Phi}(x)
    =
    U(x,0)\mathbf{\Phi}(0),
    \label{eq:landscape-spatial-system}
\end{equation}
where $\mathbf{\Phi}(x)\in\mathbb C^D$ collects
the wave-function components and their spatial
derivatives, $\mathbf M(x)$ is a bounded and
continuous matrix determined by the zero-energy
equation, and $U(x,y)$ is the evolution matrix
satisfying $U(y,y)=I_D$.

We assume that the spatial dynamics admits a
uniform exponential dichotomy on $[0,\infty)$.
At $x=0$, this gives a stable--unstable splitting
of the initial-value space:
\begin{equation}
    \mathbb C^D
    =
    \mathcal E^s\oplus\mathcal E^u.
    \label{eq:landscape-dichotomy-splitting}
\end{equation}
Here $\mathcal E^s$ is the stable subspace of
initial conditions whose solutions decay
exponentially under forward spatial evolution,
whereas $\mathcal E^u$ is a complementary
unstable subspace whose solutions grow
exponentially.

For each $y\geq0$, define
$\mathcal E^{s,u}(y)=U(y,0)\mathcal E^{s,u}$.
The uniform exponential dichotomy guarantees
constants $K\geq1$ and $\alpha>0$, independent
of $x$ and $y$, such that for all $x\geq y\geq0$,
\begin{equation}
\begin{aligned}
    \|U(x,y)v_s\|
    &\leq
    Ke^{-\alpha(x-y)}\|v_s\|,\\
    \|U(x,y)v_u\|
    &\geq
    K^{-1}e^{\alpha(x-y)}\|v_u\|,
\end{aligned}
\label{eq:landscape-dichotomy-estimates}
\end{equation}
where $v_s\in\mathcal E^s(y)$ and
$v_u\in\mathcal E^u(y)$.
The associated dichotomy projections are
uniformly bounded, so the stable and unstable
subspaces remain uniformly separated under
spatial evolution. Since $U(y,0)$ is invertible,
their dimensions are independent of $y$.

To relate this decomposition to landscape
geometry, we introduce the Hermitian quadratic
landscape
\begin{equation}
    \mathcal L
    =
    \frac12
    \mathbf{\Phi}^\dagger
    \mathbf G
    \mathbf{\Phi},
    \qquad
    \partial_x\mathcal L
    =
    \frac12
    \mathbf{\Phi}^\dagger
    \mathbf{\Gamma}_G
    \mathbf{\Phi}
    \leq0,
    \label{eq:landscape-monotonicity}
\end{equation}
where
$\mathbf{\Gamma}_G
=\partial_x\mathbf G
+\mathbf G\mathbf M
+\mathbf M^\dagger\mathbf G
\preceq0$.
An admissible landscape is specified by a
$C^1$ Hermitian matrix $\mathbf G(x)$,
with both $\mathbf G$ and $\mathbf G^{-1}$
uniformly bounded on $[0,\infty)$,
satisfying the above inequality.
Its inertia is denoted by
$\operatorname{In}\mathbf G=(n_+,n_-,0)$,
which is independent of position because
$\mathbf G(x)$ is continuous and nonsingular.

The landscape theorem established in
Ref.~\cite{qiao2026landscape} relates the inertia
of $\mathbf G$ to the stable--unstable decomposition.
The exponential decay of stable solutions,
the boundedness of $\mathbf G$, and the
monotonicity of $\mathcal L$ imply that
the quadratic form is nonnegative on
$\mathcal E^s$, yielding
$\dim\mathcal E^s\leq n_+$.
The complementary bound
$\dim\mathcal E^u\leq n_-$ follows from the
exponential growth of unstable solutions,
the uniform boundedness of $\mathbf G^{-1}$,
and the landscape monotonicity.
Since
$\dim\mathcal E^s+\dim\mathcal E^u
=n_++n_-=D$, we obtain
\begin{equation}
    n_+=\dim\mathcal E^s,
    \qquad
    n_-=\dim\mathcal E^u.
    \label{eq:landscape-inertia-dimensions}
\end{equation}

Consequently, the landscape index satisfies
\begin{equation}
    \nu_{\mathcal L}
    =
    \frac{n_+-n_-}{2}
    =
    \frac{\dim\mathcal E^s-\dim\mathcal E^u}{2}
    =
    \dim\mathcal E^s-\frac D2.
    \label{eq:landscape-inertia-index}
\end{equation}
Thus, the landscape index is determined by
the dimensions of the stable and unstable
initial-value subspaces and is independent
of the reference position and the particular
admissible landscape. The next subsection shows that the same index can also be read directly from the topology of the landscape geometry.

\subsection{Topological information encoded in landscape geometry}
\label{sec:landscape-geometry-topology}

At a fixed position $x$, define the positive and negative
regions of the quadratic landscape by
\begin{equation}
    \mathcal V_\pm(x)=
    \left\{v\in\mathbb C^D:
    \pm v^\dagger\mathbf G(x)v>0\right\}.
    \label{eq:landscape-sign-regions}
\end{equation}
These regions carry the subspace topology inherited
from $\mathbb C^D$.
We show that their topology determines the inertia
$(n_+,n_-)$ of $\mathbf G$, and hence the landscape index.

By the spectral theorem, at each fixed $x$ the
nonsingular Hermitian matrix $\mathbf G$ can be
written as
\begin{equation}
    \mathbf G
    =W
    \begin{pmatrix}
        \Lambda_+&0\\
        0&-\Lambda_-
    \end{pmatrix}
    W^\dagger,
    \label{eq:landscape-spectral-decomposition}
\end{equation}
where $W$ is unitary, while $\Lambda_+$ and
$-\Lambda_-$ contain the positive and negative
eigenvalues of $\mathbf G$, respectively.
The corresponding blocks have dimensions
$n_+$ and $n_-$ and are omitted when empty.

Introducing the transformed coordinates
\begin{equation}
    \begin{pmatrix}u\\w\end{pmatrix}
    =
    \begin{pmatrix}
        \Lambda_+^{1/2}&0\\
        0&\Lambda_-^{1/2}
    \end{pmatrix}
    W^\dagger v,
    \label{eq:landscape-canonical-coordinates}
\end{equation}
where $u\in\mathbb C^{n_+}$ and
$w\in\mathbb C^{n_-}$ correspond to the
positive and negative eigenspaces of
$\mathbf G$, respectively, the quadratic
form becomes
\begin{equation}
    v^\dagger\mathbf Gv
    =\|u\|^2-\|w\|^2.
    \label{eq:landscape-canonical-form}
\end{equation}
Since $\mathbf G$ is nonsingular,
the coordinate transformation is
invertible and defines a homeomorphism.
It maps the positive and negative
regions onto
\begin{equation}
\begin{aligned}
    \mathcal U_+
    &=\{(u,w):\|u\|>\|w\|\},\\
    \mathcal U_-
    &=\{(u,w):\|w\|>\|u\|\},
\end{aligned}
\label{eq:landscape-canonical-regions}
\end{equation}
respectively. Therefore, it can be found that $\mathcal V_\pm$ and $\mathcal U_\pm$ are homeomorphic. Below, we will show that each nonempty region \(\mathcal V_\pm\) is homotopy equivalent to \(S^{2n_\pm-1}\), where \(S^{2n_\pm-1}\) denotes the unit sphere in \(\mathbb C^{n_\pm}\), of real dimension \(2n_\pm-1\). We first consider $\mathcal U_+$ for
$n_+>0$. Since $\|u\|>\|w\|$,
the vector $u$ cannot vanish.
Let
\begin{equation}
    S_+
    =\{(u,0):\|u\|=1\}
    \subset\mathcal U_+
    \label{eq:landscape-positive-sphere}
\end{equation}
be the unit sphere in the positive
eigenspace. Define the continuous maps
\begin{equation}
\begin{aligned}
    r:\mathcal U_+&\longrightarrow S_+,
    &r(u,w)&=\left(\frac{u}{\|u\|},0\right),\\
    i:S_+&\longrightarrow\mathcal U_+,
    &i(u,0)&=(u,0),
\end{aligned}
\label{eq:landscape-positive-maps}
\end{equation}
where $r$ is a retraction and $i$
is the inclusion map.
For every point $(u,0)\in S_+$,
we have
\begin{equation}
    (r\circ i)(u,0)=(u,0),
\end{equation}
and hence
$r\circ i=\operatorname{id}_{S_+}$.

To establish the homotopy between
$i\circ r$ and the identity map
on $\mathcal U_+$, consider
\begin{equation}
    F_t^+(u,w)
    =
    \left(
    \left[1-t+\frac{t}{\|u\|}\right]u,
    (1-t)w
    \right),
    \label{eq:landscape-positive-deformation}
\end{equation}
where $0\leq t\leq1$.
Throughout this deformation,
$(u_t,w_t)=F_t^+(u,w)$ satisfies
\begin{equation}
    \|u_t\|-\|w_t\|
    =(1-t)(\|u\|-\|w\|)+t>0.
    \label{eq:landscape-positive-preservation}
\end{equation}
Thus, $F_t^+$ remains within
$\mathcal U_+$ for all $t\in[0,1]$.
At the two endpoints,
\begin{equation}
    F_0^+(u,w)=(u,w),\quad 
    F_1^+(u,w)=\left(\frac{u}{\|u\|},0\right).
\label{eq:landscape-positive-endpoints}
\end{equation}
Therefore,
\begin{equation}
    F_0^+=\operatorname{id}_{\mathcal U_+},
    \qquad
    F_1^+=i\circ r,
\end{equation}
which establishes $i\circ r\simeq\operatorname{id}_{\mathcal U_+}$.

Together with
$r\circ i=\operatorname{id}_{S_+}$,
these relations establish the
homotopy equivalence
$\mathcal U_+\simeq S_+$.
Since $u\in\mathbb C^{n_+}$,
the unit sphere $S_+$ has real
dimension $2n_+-1$.
Consequently,
\begin{equation}
    \mathcal V_+
    \simeq\mathcal U_+
    \simeq S^{2n_+-1},
    \label{eq:landscape-positive-homotopy}
\end{equation}
where $\simeq$ denotes homotopy equivalence.

The negative region $\mathcal U_-$ can be
treated analogously by exchanging $u$
and $w$. For $n_->0$, this gives
\begin{equation}
    \mathcal V_-
    \simeq\mathcal U_-
    \simeq S^{2n_--1}.
    \label{eq:landscape-negative-homotopy}
\end{equation}

Since spheres of different dimensions have distinct homotopy types, the homotopy type of each nonempty region $\mathcal V_\pm$ uniquely determines the corresponding eigenvalue count $n_\pm$. This determines the landscape inertia $(n_+,n_-)$ and hence the landscape index $\nu_L=(n_+-n_-)/2$.

By the landscape inertia theorem,
$n_+=\dim\mathcal E^s$ and
$n_-=\dim\mathcal E^u$.
The landscape index therefore connects
this geometric topology to the stable
and unstable subspaces of the zero-energy
spatial dynamics.
Its invariance under continuous
deformations is established in the
following subsection.

\subsection{Invariance under continuous deformations}
\label{sec:landscape-topological-invariance}

We now establish the invariance of the landscape
index under continuous deformations of the
Hamiltonian that preserve the uniform exponential
dichotomy. For Hamiltonians admitting a first-order
reduction with fixed dimension $D$ and uniformly
invertible highest-derivative coefficients,
continuous variations of the Hamiltonian
coefficients in the uniform norm induce continuous
deformations of $\mathbf M(x)$.

We consider the space of bounded continuous
matrix-valued functions from $[0,\infty)$
to $\mathbb C^{D\times D}$:
\begin{equation}
    \mathcal X
    =
    C_b([0,\infty),\mathbb C^{D\times D}).
    \label{eq:landscape-generator-space}
\end{equation}
Each $\mathbf M\in\mathcal X$ specifies a
first-order spatial dynamical system.
We define the distance between two elements
of $\mathcal X$ by
\begin{equation}
    d(\mathbf M_1,\mathbf M_2)
    =
    \sup_{x\geq0}
    \|\mathbf M_1(x)-\mathbf M_2(x)\|,
    \label{eq:landscape-generator-metric}
\end{equation}
which induces the topology of uniform convergence
on $\mathcal X$.

Let $\mathcal D\subset\mathcal X$ denote the subset
of matrix-valued functions whose spatial dynamics
admits a uniform exponential dichotomy on
$[0,\infty)$, as specified in the preceding
subsection. The topology of $\mathcal D$ is
inherited from $\mathcal X$.

For each $\mathbf M\in\mathcal D$, let
$\mathcal E_{\mathbf M}^s$ and
$\mathcal E_{\mathbf M}^u$ denote the stable
and unstable initial-value subspaces at $x=0$.
According to Eq.~\eqref{eq:landscape-inertia-index},
the landscape index depends only on the dimensions
of these subspaces and is independent of the
choice of admissible landscape.

For fixed $D$, the index takes values in
\begin{equation}
    \mathcal I_D
    =
    \left\{
    j-\frac D2:
    j=0,1,\ldots,D
    \right\}.
    \label{eq:landscape-index-values}
\end{equation}
Equipped with the discrete topology,
$\mathcal I_D$ is a topological space.
The landscape index defines a map
\begin{equation}
    \nu_{\mathcal L}:
    \mathcal D\longrightarrow\mathcal I_D,
    \qquad
    \mathbf M\longmapsto
    \dim\mathcal E_{\mathbf M}^s-\frac D2.
    \label{eq:landscape-index-map}
\end{equation}

The continuity of this map follows from the
roughness theorem for exponential dichotomies~\cite{Coppel1978,Ju2001}.
For every $\mathbf M_0\in\mathcal D$, there exists
$\varepsilon_0>0$ such that
$d(\mathbf M,\mathbf M_0)<\varepsilon_0$
implies $\mathbf M\in\mathcal D$ and
\begin{equation}
    \dim\mathcal E_{\mathbf M}^{s,u}
    =
    \dim\mathcal E_{\mathbf M_0}^{s,u}.
    \label{eq:landscape-roughness}
\end{equation}
Thus, $\mathcal D$ is open in $\mathcal X$,
and $\nu_{\mathcal L}$ is locally constant
on $\mathcal D$, hence continuous.

Consider a continuous deformation
$\mathbf M_\lambda\in\mathcal D$,
$0\leq\lambda\leq1$, satisfying
\begin{equation}
    \lim_{\lambda\to\lambda_0}
    d(\mathbf M_\lambda,\mathbf M_{\lambda_0})
    =
    0.
    \label{eq:landscape-admissible-path}
\end{equation}
Since $\nu_{\mathcal L}$ is continuous,
the composition
$\lambda\mapsto\nu_{\mathcal L}[\mathbf M_\lambda]$
is a continuous map from the connected interval
$[0,1]$ to the discrete space $\mathcal I_D$.
Such a map must be constant, and therefore
\begin{equation}
    \nu_{\mathcal L}[\mathbf M_\lambda]
    =
    \nu_{\mathcal L}[\mathbf M_0],
    \qquad 0\leq\lambda\leq1.
    \label{eq:landscape-topological-invariance}
\end{equation}
Thus, the landscape index is constant
on each connected component of $\mathcal D$
and is therefore a homotopy invariant
of the zero-energy spatial dynamics. Consequently, a change in the landscape index under continuous deformation requires a breakdown of the uniform exponential dichotomy.

\section{Landscape--bulk correspondence}
\label{app:roots_subspaces}

We first establish the relation between
the complex zeros of $Q(z)=\det q(z)$
and the stable and unstable subspaces
of the spatial dynamics, including the
condition for bulk gap closing.
We then use the argument principle to
derive the equivalence between the landscape
index and the winding number in the
homogeneous limit.
Finally, we compare the landscape-index difference
with the winding-number difference between two
homogeneous bulk Hamiltonians.

\subsection{Complex roots and stable--unstable subspaces}

Consider the zero-energy
equation
\begin{equation}
    q(-i\partial_x)\Psi_-(x)=0.
    \label{eq:app_q_real}
\end{equation}
In the homogeneous limit, we seek exponential solutions
$\Psi_-(x)=e^{izx}\Psi_{-,0}$, with $z\in\mathbb C$.
Substituting $-i\partial_x\to z$ into
Eq.~\eqref{eq:app_q_real} gives $q(z)\Psi_{-,0}=0$,
so a nontrivial solution requires
\begin{equation}
    Q(z)\equiv\det q(z)=0.
    \label{eq:app_Qzero}
\end{equation}
Thus, the zeros $z_j$ of $Q(z)$ determine the complex wave
numbers of the exponential zero-energy solutions.

Equation~\eqref{eq:app_q_real} can equivalently be written as
the first-order spatial system
$\partial_x\mathbf{\Phi}=\mathbf M\mathbf{\Phi}$,
where $\mathbf M$ is independent of $x$ in the homogeneous
limit and $\mathbf{\Phi}$ collects $\Psi_-$ and its spatial
derivatives. For an $N\times N$ differential operator of
order $p$ with a nonsingular highest-derivative coefficient,
the first-order system has dimension $D=pN$.
For a mode $\Psi_-(x)=e^{iz_jx}\Psi_{-,0}$, each derivative
satisfies
$\partial_x^\ell\Psi_-=(iz_j)^\ell e^{iz_jx}\Psi_{-,0}$.
Hence all components of $\mathbf{\Phi}$ carry the same
exponential factor, so that
$\mathbf{\Phi}(x)=e^{iz_jx}\mathbf{\Phi}_0$.
Substitution into the first-order equation gives
\begin{equation}
    \mathbf M\mathbf{\Phi}_0
    =
    iz_j\mathbf{\Phi}_0.
    \label{eq:app_Meigen}
\end{equation}
Thus, a zero $z_j$ of $Q(z)$ corresponds to an eigenvalue
$\lambda_j=iz_j$ of $\mathbf M$. The companion construction
preserves algebraic multiplicities, since the characteristic
polynomial of $\mathbf M$ is proportional to $Q(-i\lambda)$.

We next use this relation to identify the stable and unstable
subspaces. For a homogeneous system admitting an exponential
dichotomy~\cite{Coppel1978,Sacker1978},
the initial-value space at $x=0$ decomposes as
\begin{equation}
    \mathbb C^D
    =
    \mathcal E^s\oplus\mathcal E^u.
    \label{eq:app_dichotomy}
\end{equation}
For $x\geq0$, there exist constants $C\geq1$ and $\alpha>0$
such that
\begin{align}
    \|e^{\mathbf Mx}v_s\|
    &\leq
    Ce^{-\alpha x}\|v_s\|,
    \qquad
    v_s\in\mathcal E^s,
    \\
    \|e^{\mathbf Mx}v_u\|
    &\geq
    C^{-1}e^{\alpha x}\|v_u\|,
    \qquad
    v_u\in\mathcal E^u.
    \label{eq:app_dichotomy_bounds}
\end{align}
Thus, states in $\mathcal E^s$ decay exponentially under
forward spatial evolution, whereas those in $\mathcal E^u$
grow exponentially. 

Let $\lambda_j$ denote the distinct eigenvalues of
$\mathbf M$, and let $\mathcal G_{\lambda_j}$ be the
generalized eigenspace associated with $\lambda_j$.
The state space then decomposes as
\begin{equation}
    \mathbb C^D
    =
    \bigoplus_j\mathcal G_{\lambda_j},
    \qquad
    \dim\mathcal G_{\lambda_j}
    =
    m_{\rm alg}(\lambda_j),
    \label{eq:app_generalized_decomposition}
\end{equation}
where $m_{\rm alg}(\lambda_j)$ denotes the algebraic
multiplicity of $\lambda_j$. The generalized eigenspaces
account for repeated eigenvalues and non-diagonalizable
$\mathbf M$.

For a homogeneous system, Exponential dichotomy means that $\mathbf M$ has no purely imaginary eigenvalues. Accordingly,
\begin{equation}
    \mathcal E^s
    =
    \bigoplus_{\operatorname{Re}\lambda_j<0}
    \mathcal G_{\lambda_j},
    \qquad
    \mathcal E^u
    =
    \bigoplus_{\operatorname{Re}\lambda_j>0}
    \mathcal G_{\lambda_j}.
    \label{eq:app_stable_unstable}
\end{equation}
Their dimensions are therefore
\begin{equation}
    \dim\mathcal E^s
    =
    \sum_{\operatorname{Re}\lambda_j<0}
    m_{\rm alg}(\lambda_j),
    \quad
    \dim\mathcal E^u
    =
    \sum_{\operatorname{Re}\lambda_j>0}
    m_{\rm alg}(\lambda_j).
    \label{eq:app_dimensions}
\end{equation}
Since $\lambda_j=iz_j$, we have
$\operatorname{Re}\lambda_j=-\operatorname{Im}z_j$.
Zeros of $Q(z)$ in the upper half-plane therefore
correspond to the generalized eigenspaces contributing
to $\mathcal E^s$, whereas zeros in the lower half-plane
correspond to those contributing to $\mathcal E^u$.
Counting algebraic multiplicities, the numbers $N_+$
and $N_-$ of zeros in the upper and lower half-planes
satisfy
\begin{equation}
    N_+
    =
    \dim\mathcal E^s,
    \qquad
    N_-
    =
    \dim\mathcal E^u.
    \label{eq:app_roots_subspaces}
\end{equation}

For a homogeneous system, $\mathbf M(x)=\mathbf M$,
the exponential dichotomy fails precisely when
the spectrum of $\mathbf M$ intersects the
imaginary axis. Using the relation
$\lambda_{\rm sp}=iz$ established in Appendix~\ref{app:roots_subspaces},
\begin{equation}
    \sigma(\mathbf M)\cap i\mathbb R
    \neq\varnothing
    \quad\Longleftrightarrow\quad
    \exists\,k\in\mathbb R:
    Q(k)=0.
    \label{eq:landscape-homogeneous-critical-condition}
\end{equation}
This is the zero-energy bulk-gap-closing condition.

\subsection{Contour representation of the winding number}

For the finite-order homogeneous systems considered here,
$Q(z)$ is a degree-$D$ polynomial.
A gapped bulk satisfies $Q(k)\neq0$ for every real $k$.
The continuum winding number defined in
Eq.~\eqref{eq:winding} is
\begin{equation}
    \nu_c
    =\frac{1}{2\pi i}
      \int_{-\infty}^{\infty}
      dk\,\partial_k\ln Q(k).
    \label{eq:app_winding_Q}
\end{equation}

To evaluate Eq.~\eqref{eq:app_winding_Q}, consider the
positively oriented contour formed by the interval
$[-R,R]$ and the upper semicircle $C_R$.
Since $Q(z)$ is a polynomial, it has no finite poles
and only finitely many zeros. For sufficiently large $R$,
the contour encloses all zeros of $Q(z)$ in the upper
half-plane. The argument principle then gives
\begin{equation}
    \frac{1}{2\pi i}
    \oint
    dz\,
    \frac{Q'(z)}{Q(z)}
    =
    N_+,
    \label{eq:app_argument_principle}
\end{equation}
where the zeros, including repeated ones, are counted
with algebraic multiplicity.

Separating the real-axis and semicircle contributions yields
\begin{equation}
    \frac{1}{2\pi i}
    \int_{-R}^{R}
    dk\,
    \frac{Q'(k)}{Q(k)}
    =
    N_+
    -
    \frac{1}{2\pi i}
    \int_{C_R}
    dz\,
    \frac{Q'(z)}{Q(z)}.
    \label{eq:app_contour_split}
\end{equation}
For large $|z|$, a degree-$D$ polynomial satisfies
\begin{equation}
    Q(z)
    =
    c_D z^D
    \left[
        1+\mathcal O(z^{-1})
    \right],
    \qquad
    c_D\neq0,
    \label{eq:app_Q_asymptotic}
\end{equation}
and hence
\begin{equation}
    \frac{Q'(z)}{Q(z)}
    =
    \frac{D}{z}
    +
    \mathcal O(z^{-2}).
    \label{eq:app_log_derivative_asymptotic}
\end{equation}
Parameterizing the upper semicircle as
$z=Re^{i\theta}$ with $0\leq\theta\leq\pi$ gives
\begin{align}
    \frac{1}{2\pi i}
    \int_{C_R}
    dz\,
    \frac{Q'(z)}{Q(z)}
    &=
    \frac{1}{2\pi i}
    \int_0^\pi
    iRe^{i\theta}
    \left[
        \frac{D}{Re^{i\theta}}
        +
        \mathcal O(R^{-2})
    \right]
    d\theta
    \nonumber\\
    &=
    \frac{D}{2}
    +
    \mathcal O(R^{-1}).
    \label{eq:app_semicircle}
\end{align}
Taking $R\to\infty$ therefore gives
\begin{equation}
    \nu_c
    =
    N_+
    -
    \frac{D}{2}.
    \label{eq:app_nu_Nplus}
\end{equation}

By the fundamental theorem of algebra, a degree-$D$
polynomial has exactly $D$ zeros in the complex plane
when algebraic multiplicities are counted.
Since $Q(k)\neq0$ for all real $k$, no zero lies on
the real axis, and therefore
\begin{equation}
    N_+ + N_- = D.
    \label{eq:app_root_sum}
\end{equation}
Combining Eqs.~\eqref{eq:app_nu_Nplus} and
\eqref{eq:app_root_sum}, we obtain
\begin{equation}
    \nu_c
    =
    \frac{N_+-N_-}{2}.
    \label{eq:app_winding_roots}
\end{equation}

Using the correspondence established in Appendix~\ref{app:roots_subspaces},
$N_+=\dim\mathcal E^s$ and $N_-=\dim\mathcal E^u$,
this becomes
\begin{equation}
   \nu_c
    =
    \frac{\dim\mathcal E^s-\dim\mathcal E^u}{2}
    =
    \nu_{\mathcal L}.
    \label{eq:app_bulk_correspondence}
\end{equation}
Thus, in the homogeneous limit, the landscape index
coincides with the winding number of the continuum Hamiltonian.

The relation holds for both even and odd $D$.
Since $\nu_c=N_+-D/2$, the individual winding numbers and landscape
indices are integers for even $D$ and half-integers for odd $D$.
Their differences at fixed $D$ are always integers, as shown next.

\subsection{Relative winding numbers}

Consider two homogeneous bulk Hamiltonians, labeled $L$ and $R$,
with the same spatial-system dimension $D$.
Let $Q^{L,R}(z)$ be their characteristic polynomials and
$\nu_c^{L,R}$ and $\nu_{\mathcal L}^{L,R}$ their continuum winding
numbers and landscape indices.
Both bulks are gapped, so $Q^{L,R}(k)\neq0$ for all real $k$.
Since both polynomials have degree $D$,
\begin{equation}
    \frac{Q^R(z)}{Q^L(z)}
    =\frac{c_D^R}{c_D^L}\left[1+\mathcal O(z^{-1})\right],
    \qquad |z|\to\infty,
    \label{eq:app_ratio_asymptotic}
\end{equation}
where $c_D^L$ and $c_D^R$ are the nonzero leading coefficients of $\mathcal Q^L(z)$ and $\mathcal Q^R(z)$, respectively.
The ratio approaches the same nonzero value at
$k\to-\infty$ and $k\to+\infty$, so its image forms a closed loop
in the complex plane.
Moreover,
$\partial_z\ln[Q^R(z)/Q^L(z)]=\mathcal O(z^{-2})$,
and the upper-semicircle contribution vanishes as $R\to\infty$.

Applying the argument principle to this ratio gives
\begin{align}
    \nu_c^R-\nu_c^L
    &=\frac{1}{2\pi i}
      \int_{-\infty}^{\infty}dk\,
      \partial_k\ln\frac{Q^R(k)}{Q^L(k)}
      \nonumber\\
    &=N_+^R-N_+^L\in\mathbb Z.
    \label{eq:app_winding_difference}
\end{align}
Here, zeros of $Q^R$ and $Q^L$ contribute as zeros and poles,
respectively, with algebraic multiplicities.
Using $\nu_c^{L,R}=\nu_{\mathcal L}^{L,R}$, we obtain
\begin{equation}
    \begin{aligned}
        \nu_{\mathcal L}^{R}-\nu_{\mathcal L}^{L}
        &=\nu_c^{R}-\nu_c^{L}\\
        &=N_+^{R}-N_+^{L}\in\mathbb Z.
    \end{aligned}
    \label{eq:app_index_difference}
\end{equation}
The landscape-index difference therefore equals the bulk
winding-number difference and remains integer-valued even when the
individual indices are half-integers.

\section{Landscape--boundary correspondence}
\label{SM:Fredholm}
We now prove that the difference between the landscape
indices on the two sides of an interface equals the
net chiral zero-mode count, establishing the
landscape--boundary correspondence. To this end, we first
express the zero-mode count in each chiral sector in
terms of its decaying subspaces and then use the duality
between the two chiral spatial systems to show that
the net count depends only on the dimensions of
these subspaces. 

\subsection{Interface zero modes and decaying subspaces}

Consider an interface at $x=0$ between two
spatially inhomogeneous chiral regions.
We assume that the zero-energy spatial
dynamics admits uniform exponential
dichotomies on both half-lines.

The zero-energy equations in the two
chiral sectors are
\begin{equation}
    q\Psi_-=0,
    \qquad
    q^\dagger\Psi_+=0,
    \label{SM:chiral-equations}
\end{equation}
which can be recast as first-order
spatial systems,
\begin{equation}
    \partial_x\mathbf\Phi_\pm
    =
    \mathbf M_\pm(x)\mathbf\Phi_\pm,
    \label{SM:chiral-spatial-systems}
\end{equation}
where $\mathbf\Phi_\pm$ contain the wave
functions and their spatial derivatives.

For $q\Psi_-=0$, let
$\mathcal E_L^u(0)$ and
$\mathcal E_R^s(0)$ denote the
subspaces of initial conditions
whose solutions decay as
$x\to-\infty$ and $x\to+\infty$,
respectively.
A localized zero mode must decay
in both directions and satisfy
$\mathbf\Phi_-(0^-)=\mathbf\Phi_-(0^+)$.
Its number is therefore
\begin{equation}
    \mathcal N_-
    =
    \dim\ker q
    =
    \dim\left(
    \mathcal E_L^u(0)
    \cap
    \mathcal E_R^s(0)
    \right).
    \label{SM:kernel}
\end{equation}

Similarly, for $q^\dagger\Psi_+=0$,
let $\mathcal F_L^u(0)$ and
$\mathcal F_R^s(0)$ denote the
corresponding decaying subspaces.
Applying the same matching condition gives
\begin{equation}
    \mathcal N_+
    =
    \dim\ker q^\dagger
    =
    \dim\left(
    \mathcal F_L^u(0)
    \cap
    \mathcal F_R^s(0)
    \right).
    \label{SM:adjoint-kernel}
\end{equation}
The net chiral zero-mode count is
$\Delta\mathcal N
=\mathcal N_- -\mathcal N_+$.

\subsection{Duality of chiral spatial dynamics}

To relate the decaying subspaces of the two
chiral sectors, we use the Lagrange identity
to transform the spatial system for
$q^\dagger$ into the dual system of that
for $q$. The transformation and its inverse
are uniformly bounded, so exponential
decay is preserved.

Consider the general finite-order
differential operator
\begin{equation}
    q(x,-i\partial_x)
    =
    \sum_{j=0}^{p}
    q_j(x)(-i\partial_x)^j,
    \qquad p\geq1,
    \label{SM:general-operator}
\end{equation}
Here, $q_j(x)$ are $N\times N$ matrix functions
that are continuously differentiable up to
order $p$ on the whole real line. The matrices
and these derivatives are uniformly bounded,
and the highest-derivative coefficient
$q_p(x)$ is uniformly invertible:
\begin{equation}
    \sup_{x\in\mathbb R}
    \|q_p^{-1}(x)\|<\infty.
    \label{SM:principal}
\end{equation}
The corresponding spatial system
has dimension $D=pN$.

The adjoint equation takes the form
\begin{equation}
    q^\dagger\Psi_+
    =
    \sum_{j=0}^{p}
    (-i\partial_x)^j
    \left[q_j^\dagger(x)\Psi_+\right]
    =0,
    \label{SM:adjoint-operator}
\end{equation}
where the derivatives act on both
the coefficients and the wave function.

We introduce the transformed components
\begin{equation}
    \widetilde{\Phi}_{+,l}
    =
    i^{l+1}
    \sum_{j=l+1}^{p}
    (-i\partial_x)^{j-l-1}
    \left[q_j^\dagger(x)\Psi_+\right],
    \label{SM:dual-components}
\end{equation}
with $l=0,\ldots,p-1$.
Together, they form
$\widetilde{\mathbf\Phi}_+\in\mathbb C^D$
and define the linear transformation
\begin{equation}
    \widetilde{\mathbf\Phi}_+
    =
    \mathbf T(x)\mathbf\Phi_+.
    \label{SM:dual-transformation}
\end{equation}

Here, $\mathbf\Phi_+$ collect $\Psi_\pm$ and their spatial derivatives up to order $p-1$, following the first-order reduction introduced after Eq.~(1). Expanding the derivatives in
Eq.~\eqref{SM:dual-components},
for $p=2$ we obtain
\begin{equation}
    \mathbf T(x)
    =
    \begin{pmatrix}
        iq_1^\dagger+\partial_xq_2^\dagger
        &q_2^\dagger\\
        -q_2^\dagger&0
    \end{pmatrix}.
    \label{SM:dual-transformation-example}
\end{equation}
For general $p$, $\mathbf T(x)$
is block anti-triangular, with
anti-diagonal blocks
\begin{equation}
    \mathbf T_{l,p-1-l}(x)
    =
    i^{l+1}(-i)^{p-l-1}q_p^\dagger(x).
    \label{SM:dual-antidiagonal}
\end{equation}
Since these blocks are invertible
and the coefficients and their
required derivatives are bounded,
both $\mathbf T(x)$ and
$\mathbf T^{-1}(x)$ are
uniformly bounded.

We next verify the transformed
spatial dynamics.
Using Eq.~\eqref{SM:adjoint-operator},
differentiation of
Eq.~\eqref{SM:dual-components} gives
\begin{equation}
    \partial_x\widetilde{\Phi}_{+,0}
    =
    -\sum_{j=1}^{p}
    (-i\partial_x)^j
    [q_j^\dagger\Psi_+]
    =
    q_0^\dagger\Psi_+.
    \label{SM:dual-first-component}
\end{equation}
For $l=1,\ldots,p-1$,
we similarly obtain
\begin{equation}
\begin{aligned}
    \partial_x\widetilde{\Phi}_{+,l}
    &=
    -i^l
    \sum_{j=l+1}^{p}
    (-i\partial_x)^{j-l}
    [q_j^\dagger\Psi_+]\\
    &=
    -\widetilde{\Phi}_{+,l-1}
    +i^lq_l^\dagger\Psi_+.
\end{aligned}
\label{SM:dual-component-equations}
\end{equation}

On the other hand, the matrix
$\mathbf M_-$ associated with
$q\Psi_-=0$ is
\begin{equation}
    \mathbf M_-
    =
    \begin{pmatrix}
        0&I_N&0&\cdots&0\\
        0&0&I_N&\cdots&0\\
        \vdots&\vdots&\vdots&\ddots&\vdots\\
        0&0&0&\cdots&I_N\\
        C_0&C_1&C_2&\cdots&C_{p-1}
    \end{pmatrix},
    \label{SM:companion-matrix}
\end{equation}
where $I_N$ is the identity matrix and
\begin{equation}
    C_l
    =
    -(-i)^{l-p}q_p^{-1}q_l,
    \qquad l=0,\ldots,p-1.
    \label{SM:companion-blocks}
\end{equation}
The last row follows from solving
$q\Psi_-=0$ for its highest derivative.

Taking the negative Hermitian conjugate
of $\mathbf M_-$ and applying it to
$\widetilde{\mathbf\Phi}_+$,
the identity blocks contribute
$-\widetilde{\Phi}_{+,l-1}$,
while the remaining blocks give
$i^lq_l^\dagger\Psi_+$, since
$\widetilde{\Phi}_{+,p-1}
=i^pq_p^\dagger\Psi_+$.
These terms reproduce
Eqs.~\eqref{SM:dual-first-component}
and \eqref{SM:dual-component-equations},
yielding
\begin{equation}
    \partial_x\widetilde{\mathbf\Phi}_+
    =
    -\mathbf M_-^\dagger(x)
    \widetilde{\mathbf\Phi}_+.
    \label{SM:dual-system}
\end{equation}
Thus, the positive-chirality spatial
dynamics is related to the dual
of the negative-chirality system
through an invertible transformation.

Combining Eq.~\eqref{SM:dual-system}
with $\partial_x\mathbf\Phi_-=
\mathbf M_-\mathbf\Phi_-$ gives
\begin{equation}
    \partial_x
    \left(
    \widetilde{\mathbf\Phi}_+^\dagger
    \mathbf\Phi_-
    \right)
    =0.
    \label{SM:conserved-pairing}
\end{equation}
The inner product is therefore
constant along the spatial evolution.

For solutions decaying toward
the same spatial infinity,
this constant must vanish.
The decaying subspaces of the
dual systems also have
complementary dimensions.
Consequently,
\begin{equation}
\begin{aligned}
    \mathbf T(0)\mathcal F_L^u(0)
    &=
    \left(\mathcal E_L^u(0)\right)^\perp,\\
    \mathbf T(0)\mathcal F_R^s(0)
    &=
    \left(\mathcal E_R^s(0)\right)^\perp.
\end{aligned}
\label{SM:dual-subspaces}
\end{equation}
Here the orthogonal complements
are taken in $\mathbb C^D$.

Since $\mathbf T(0)$ is invertible,
Eqs.~\eqref{SM:adjoint-kernel}
and \eqref{SM:dual-subspaces} give
\begin{equation}
\begin{aligned}
    \mathcal N_+
    &=
    \dim\left[
    \left(\mathcal E_L^u\right)^\perp
    \cap
    \left(\mathcal E_R^s\right)^\perp
    \right]\\
    &=
    D-\dim\left(
    \mathcal E_L^u+
    \mathcal E_R^s
    \right),
\end{aligned}
\label{SM:positive-zero-mode-count}
\end{equation}
where all subspaces are evaluated
at $x=0$, and we used
$(A+B)^\perp=A^\perp\cap B^\perp$.

Combining this result with
Eq.~\eqref{SM:kernel} and using
$\dim(A\cap B)+\dim(A+B)
=\dim A+\dim B$, we obtain
\begin{equation}
\begin{aligned}
    \Delta\mathcal N
    &=
    \mathcal N_- -\mathcal N_+\\
    &=
    \dim\mathcal E_L^u
    +\dim\mathcal E_R^s-D.
\end{aligned}
\label{SM:index-spatial}
\end{equation}

Finally, the landscape geometry
relates these subspace dimensions
to the inertia of the landscape
matrix:
\begin{equation}
    \dim\mathcal E_L^u=n_-^L,
    \qquad
    \dim\mathcal E_R^s=n_+^R.
    \label{SM:landscape-subspace-dimensions}
\end{equation}
Substituting these dimensions into
Eq.~\eqref{SM:index-spatial} gives
\begin{equation}
    \Delta\mathcal N
    =n_-^L+n_+^R-D
    =\nu_{\mathcal L}^{R}-\nu_{\mathcal L}^{L},
\end{equation}
which establishes the landscape--boundary correspondence.

This landscape--boundary correspondence
applies to spatially inhomogeneous systems
admitting uniform exponential dichotomies
on both half-lines, whereas asymptotically
homogeneous systems constitute only
a special case.

\section{Analytical bounds for modulated QAH systems}
\label{SM:dimensional-reduction}

This appendix derives analytical index bounds for
periodically modulated QAH systems and verifies
the required exponential dichotomy using spatial
Floquet theory. It then establishes a
profile-independent interface criterion from
bounds on the mass and its spatial gradient.

\subsection{Periodic mass modulation and spatial Floquet analysis}
\label{SM:QAH}

For the QAH Hamiltonian in Eq.~\eqref{eq:qah-model}, consider
\begin{equation}
    m(x)=m_0+\delta m\sin(x/\ell),
    \qquad \delta m>0,\quad\ell>0.
    \label{SM:QAH-periodic-mass}
\end{equation}
The period is $T=2\pi\ell$. At fixed $k_y$, define
$m_{\rm eff}(k_y)=m_0+Bk_y^2$, so that
\begin{equation}
    \mu(x;k_y)=-\frac{m_{\rm eff}(k_y)+\delta m\sin(x/\ell)}{B}.
    \label{SM:QAH-periodic-mu}
\end{equation}
For $|m_{\rm eff}|>\delta m$, the effective parameter has a fixed
nonzero sign. Its logarithmic gradient is
\begin{equation}
    \left|\partial_x\ln|\mu|\right|
    =\frac{\delta m|\cos(x/\ell)|}
           {\ell|m_{\rm eff}+\delta m\sin(x/\ell)|}.
\end{equation}
Maximizing over a period gives the stationary condition
$\sin(x/\ell)=-\delta m/m_{\rm eff}$ and the exact maximum
\begin{equation}
    \sup_x\left|\partial_x\ln|\mu(x;k_y)|\right|
    =\frac{\delta m}{\ell\sqrt{m_{\rm eff}^2-\delta m^2}}.
    \label{SM:QAH-periodic-gradient}
\end{equation}
The sufficient condition in Eq.~\eqref{SM:QAH-gradient-condition}
therefore becomes
\begin{equation}
    |m_{\rm eff}|>\delta m C,
    \qquad
    C=\sqrt{1+\frac{B^2}{4A^2\ell^2}}.
    \label{SM:QAH-periodic-condition}
\end{equation}
Since $C>1$, this also ensures that $\mu$ is separated from zero.
The sign of $\mu$ fixes the landscape inertia, yielding
\begin{equation}
    \nu_{\mathcal L}(k_y)=
    \begin{cases}
        1,&m_0+Bk_y^2<-\delta m C,\\
        0,&m_0+Bk_y^2>\delta m C.
    \end{cases}
    \label{SM:QAH-periodic-index}
\end{equation}
For $m_0<-\delta m C$, the hard-wall problem supports one protected
chiral edge state with $E(k_y)=-Ak_y$ within
\begin{equation}
    k_y^2<-\frac{m_0+\delta m C}{B}.
    \label{SM:QAH-periodic-window}
\end{equation}
For $m_0>\delta m C$, all momentum channels have zero index. These inequalities give sufficient analytical bounds for the sinusoidal family. 

For an independent calculation of the channel index,
introduce the evolution matrix $\mathbf U(x)$ through
$\mathbf\Phi(x)=\mathbf U(x)\mathbf\Phi(0)$,
where $\mathbf\Phi=(\phi,\phi')^T$.
It satisfies
\begin{equation}
    \partial_x\mathbf U=\mathbf M(x;k_y)\mathbf U,
    \qquad \mathbf U(0)=\mathbb I_2,
    \label{SM:QAH-spatial}
\end{equation}
where
\begin{equation}
    \mathbf M(x;k_y)=
    \begin{bmatrix}
        0&1\\
        -\mu(x;k_y)&-A/B
    \end{bmatrix}.
\end{equation}
The one-period evolution matrix is $\mathbf P=\mathbf U(T)$. If $\rho_j$ are its eigenvalues, their real growth rates are
\begin{equation}
    \lambda_j=\frac{1}{T}\ln|\rho_j|.
    \label{SM:QAH-Floquet-exponents}
\end{equation}
A multiplier with $|\rho_j|<1$ corresponds to a
solution that decays over successive spatial periods.
Since $\mathbf U(0)=\mathbb I_2$ and
$\operatorname{tr}\mathbf M=-A/B$,
Liouville's formula gives
\begin{equation}
        \det\mathbf P
        =\exp\left[
            \int_0^T\operatorname{tr}\mathbf M(x;k_y)\,dx
          \right]
        =e^{-(A/B)T}.
    \label{SM:QAH-Liouville}
\end{equation}
The determinant is also the product of the two
Floquet multipliers, $\rho_1\rho_2=\det\mathbf P$.
Using $\lambda_j=T^{-1}\ln|\rho_j|$, we obtain
\begin{equation}
    \lambda_1+\lambda_2
    =\frac{1}{T}\ln|\det\mathbf P|
    =-\frac{A}{B}.
\end{equation}
Thus, a negative largest exponent means that both independent solutions
decay and $\nu_{\mathcal L}=1$. A positive largest exponent means
one growing and one decaying solution, giving $\nu_{\mathcal L}=0$.
A multiplier on the unit circle signals a breakdown of the exponential
dichotomy.

The analytical bounds also guarantee the exponential
dichotomy required for the landscape index.
Consider the periodic landscape
$\mathcal L=(|\phi'|^2+\mu|\phi|^2)/(2|\mu|)$. For a nonzero Floquet solution,
$\mathbf\Phi(T)=\rho\mathbf\Phi(0)$.
The periodicity $\mu(T;k_y)=\mu(0;k_y)$ therefore
implies $\mathcal L(T)=|\rho|^2\mathcal L(0)$.
Consequently, 
\begin{equation}
    \begin{aligned}
        (|\rho|^2-1)\mathcal L(0)
        &=\mathcal L(T)-\mathcal L(0)\\
        &=-\int_0^T
          \frac{A/B+\tfrac12\partial_x\ln|\mu|}{|\mu|}
          |\phi'|^2\,dx<0.
    \end{aligned}
    \label{SM:QAH-Floquet}
\end{equation}
The strict inequality follows because the coefficient
in the integral is uniformly positive and a nonzero
solution cannot have $\phi'\equiv0$ over a period when $\mu\neq0$.
If $|\rho|=1$, the landscape would return to its
initial value after one period, contradicting this
strict decrease.
Thus, all Floquet multipliers satisfy $|\rho|\neq1$,
establishing a uniform exponential
dichotomy~\cite{DaleckiiKrein1974}.

\subsection{Interface bounds and protected chiral modes}
\label{SM:QAH-interface}

The interface criterion applies to any smooth bounded
mass profile satisfying
\begin{equation}
    \begin{alignedat}{2}
        m(x)&\geq m_*,   &\quad x&\leq-x_*,\\
        m(x)&\leq-m_*,   &\quad x&\geq x_*,\\
        |m'(x)|&\leq b_*,&\quad |x|&\geq x_*,
    \end{alignedat}
    \label{SM:QAH-tail-bounds}
\end{equation}
where $x_*>0$, $m_*>0$, and $b_*\geq0$.
For $Bk_y^2<m_*$, the effective parameter is
negative on the left tail and positive on the right,
with
\begin{equation}
    |\mu(x;k_y)|\geq
    \begin{cases}
        \dfrac{m_*}{B}+k_y^2,&x\leq-x_*,\\[2mm]
        \dfrac{m_*}{B}-k_y^2,&x\geq x_*.
    \end{cases}
\end{equation}
Since $|\partial_x\mu|\leq b_*/B$ on both tails,
a common upper bound is
\begin{equation}
    \left|\partial_x\ln|\mu(x;k_y)|\right|
    \leq\frac{b_*}{m_*-Bk_y^2}.
\end{equation}
Equation~\eqref{SM:QAH-gradient-condition} therefore
holds with a uniform positive margin for each
fixed momentum satisfying
\begin{equation}
    k_y^2<\frac{m_*}{B}-\frac{b_*}{2A}.
    \label{SM:QAH-interface-window}
\end{equation}
The strict Lyapunov argument used in the main
text establishes exponential dichotomies on
both tails under these bounds. The tail indices are consequently
$\nu_{\mathcal L}^{L}=0$ and
$\nu_{\mathcal L}^{R}=1$.
Propagation through the finite central region is
invertible and preserves the dimensions of the
decaying subspaces.
No fixed-sign condition on $\mu$ is required
in this central region.
At the interface, $\dim\mathcal E_L^u=1$ and
$\dim\mathcal E_R^s=2$, giving
\begin{equation}
    \begin{aligned}
        \mathcal N_-
        &=\dim\bigl(\mathcal E_L^u\cap\mathcal E_R^s\bigr)=1,\\
        \mathcal N_+
        &=2-\dim\bigl(\mathcal E_L^u+\mathcal E_R^s\bigr)=0.
    \end{aligned}
    \label{SM:QAH-interface-count}
\end{equation}
The landscape--boundary correspondence thus
guarantees one protected negative-chirality
interface branch with $E(k_y)=-Ak_y$ for every
smooth bounded mass profile satisfying these
tail bounds.


\bibliography{mainRef}

\end{document}